\documentclass[aps,prb,twocolumn,groupedaddress,superscriptaddress]{revtex4}

\usepackage[dvips]{graphicx}
\usepackage{epstopdf}
\usepackage{amssymb}
\usepackage{enumerate}

\usepackage{mathrsfs}
\usepackage{dcolumn}
\usepackage{bm}
\usepackage{color}

\begin{document}

\title{ Multifractality of random eigenfunctions and generalization of Jarzynski equality}

\author{  I.~M.~Khaymovich}
 \email{ivan.khaymovich@aalto.fi}
 \affiliation{Low Temperature Laboratory, O.V. Lounasmaa Laboratory, Aalto University, FI-00076 Aalto, Finland}
 \affiliation{Institute for Physics of Microstructures, Russian Academy of Sciences, 603950 Nizhny Novgorod, GSP-105, Russia }
 \author{ J.~V.~Koski}
 \affiliation{Low Temperature Laboratory, O.V. Lounasmaa Laboratory, Aalto University, FI-00076 Aalto, Finland}
 \author{ O.-P.~Saira}
 \affiliation{Low Temperature Laboratory, O.V. Lounasmaa Laboratory, Aalto University, FI-00076 Aalto, Finland}
 \author{V.~E.~Kravtsov}
  \affiliation{Abdus Salam International Center for Theoretical Physics, Strada Costiera 11, 34151 Trieste, Italy}
 \affiliation{L. D. Landau Institute for Theoretical Physics, Chernogolovka, Russia}
\author{J.~P.~Pekola}
 \affiliation{Low Temperature Laboratory, O.V. Lounasmaa Laboratory, Aalto University, FI-00076 Aalto, Finland}
\begin{abstract}

{\bf  Systems driven out of equilibrium experience large fluctuations of the dissipated work. The same is true for wave function amplitudes in disordered systems close to the Anderson localization transition.  In both cases the probability distribution function is given by the large deviation ansatz.
Here we exploit the analogy between the statistics of work dissipated in a driven single-electron box and that of  random multifractal wave function amplitudes  and uncover  new relations which generalize the Jarzynski equality.
We checked the new relations theoretically using the rate equations for sequential tunneling of electrons  and experimentally by measuring the dissipated work in a driven  single-electron box and found a remarkable correspondence. The results represent an important universal feature of the work statistics in systems out of equilibrium and help to understand the nature of the symmetry of multifractal exponents   in the theory of Anderson localization.}
\end{abstract}

\maketitle

\section{Introduction}
Unlike the adiabatic processes where the work $W$ done on the system is equal to the difference in the free energy $\Delta F$, the non-adiabatic drive protocols are associated with work that depends not only on the parameters of the system and details of the drive protocol but also experiences fluctuations relative to its average value.\cite{RNA_JE_test, Bead_in_water_Crooks_test, RNA_Crooks_test, Two-levels_in_Di_JE_test, RNA_JE_test_Ritort, SEB_JE_test_Saira,Japanese_QDots,Kung_QDots}
Statistics of work can be described by the probability distribution function (PDF), $P_{w}(W)$, and it is an important goal to find universal features in $P_{w}(W)$ that remain unchanged within certain universality classes.\cite{Silva}
The best known relations of this kind are the Jarzynski equality \cite{BKE-1,BKE-2, JE} and the Crooks relation.\cite{Crooks}
The former one states that the exponent $e^{-(W - \Delta F)/k_{\rm B}T}$ averaged over repeated identical driving protocols is equal to 1, where $T$ is the temperature of the single bath 
and $k_{\rm B}$ is the Boltzmann constant. This necessarily implies that during some drive realizations the dissipated work $W - \Delta F$ must be negative in a naive (and wrong) contradiction with the Second Law which only states that the average dissipated work remains positive. The Crooks relation
\begin{equation}\label{Crooks}
\frac{P_w(W)}{\tilde P_w(-W)} = e^{(W-\Delta F) / k_{\rm B}T}
\end{equation}
concerns the PDF's of work in the direct ($P_{w}(W)$) and time-reversed ($\tilde{P}_{w}(-W)$) processes. This relation has many important consequences (with the Jarzynski equality being one of them) and practical applications, e.g. in the determination of free energy of folding proteins.\cite{RNA_JE_test,RNA_JE_test_Ritort}

We use the Crooks relation to find a correspondence between statistics of work in a broad class of systems driven by time-reversal symmetric protocols and statistics of random multifractal wave functions in disordered quantum systems close to the Anderson localization transition.\cite{Mirlin-review, KrMutta, RMT-book} The unifying principle of this correspondence \cite{Monthus} is the so called Large Deviation Principle~\cite{Touchette} according to which the PDF of a large variety of systems takes the form of the large deviation ansatz (LDA),
\begin{equation}\label{LDA}
P_{{\rm LDA}}(S)\sim {\rm exp}\left[-\mathfrak{n}\,G(S/\mathfrak{n}) \right],\;\;\;\;\;\mathfrak{n}\gg 1,
\end{equation}
where $G(y)$ is a system-specific function.
The LDA can be viewed as a generalization of the Central Limit Theorem of statistics according to which the sum $S$ of a large number $\mathfrak{n}$ of identically distributed independently fluctuating quantities $s_{k}$ has a limiting Gaussian distribution with the variance $\sigma^2\propto \mathfrak{n}$. Indeed, if we require that  $G(y)$ in Eq.~(\ref{LDA}) has a minimum, the expansion of this function near this minimum immediately results in the correct Gaussian PDF. The significance of the LDA is that it also describes the non-Gaussian tails of the distribution. Different realizations of LDA are characterized by different functions $G(y)$ and different effective number $\mathfrak{n}$ of independently fluctuating quantities.
For example, in the discrete Markov process (or Markov chain) driving time $t$ plays the role of large parameter $\mathfrak n$ for steady state distributions of dissipated work \cite{Esposito_PRE88} and heat \cite{Fogedby_Imparato_2011}.

Critical eigenfunctions $\psi_{i}$ ($i=1,...N$) near the Anderson localization transition and in certain random matrix ensembles have  multifractal statistics.\cite{Mirlin-review, KrMutta, RMT-book} A characteristic feature of such  statistics is that the eigenfunction amplitude $|\psi_{i}|^{2}$ takes a broad set of values (at different sites $i$ or in different realizations of disorder) which scale like $|\psi_{i}|^{2}\sim N^{-\alpha}$ ($\alpha>0$) with the total number of sites $N$   in a disordered tight-binding lattice (or the matrix size). The number of sites on a lattice where scaling is characterized by a certain $\alpha$ is $M\sim N^{f(\alpha)}$, where $f(\alpha)$ is known as the spectrum of fractal dimensions.
Were $\alpha$ taking only one single value $\alpha_{0}$, the set of ``occupied'' sites on the $d$-dimensional lattice would be a fractal with the Hausdorff dimension $d_{h}=d f(\alpha_{0})$.
Multifractality implies that there is a range of possible values of $\alpha$ with the corresponding range of fractal dimensions $f(\alpha)$.
In the language of LDA this implies that PDF  of the amplitude $|\psi_{i}|^{2}$ has a form Eq.~(\ref{LDA}) with
$S=-\ln(N|\psi|^{2})$ and $\mathfrak{n}=\ln N$.  The function $G(y)$ is
related with the multifractality spectrum $f(\alpha)$ as
 $G(y)=1-f(1+y)$.\cite{Mirlin-review}
It depends on parameters of the system such as the dimensionality or the bandwidth of the random matrix ensemble  and has a non-trivial limit at $N\rightarrow\infty$. There is a remarkable symmetry,\cite{Monthus,MF_q_1-q_symmetry, Mirlin-review}
\begin{equation}\label{MF-sym}
f(1+y) = f(1-y)+y,\quad \Leftrightarrow\quad G(y)=G(-y)-y \ ,
\end{equation}
whose physical origin is perhaps deeper than a current understanding \cite{Fyodorov-Savin} based on ``full chaotization'' of particle dynamics in a random potential which leads to the homogeneous distribution of the scattering phase off the disordered system.

An important observation\cite{Monthus}
with potentially very far-reaching consequences is that  within the LDA the symmetry Eq.~(\ref{MF-sym})
is equivalent to the Crooks-like relation,
\begin{equation}\label{CrooksMF}
\frac{P_{{\rm LDA}}(y)}{P_{{\rm LDA}}(-y)}=e^{\mathfrak{n}\,y}, \quad y=S/\mathfrak{n} \ .
\end{equation}
In this work we formulate   a generalization, Eq.~(\ref{generalization}), of the Jarzynski equality for the work generating function. This generalization has been proven theoretically by a stochastic calculus using the rate equations and experimentally for a driven single electron box (SEB) in the Coulomb blockade regime.
\section{Results}
\subsection{The large-deviation parameter and the temperature}
To formulate a dictionary between the statistics of work in driven systems and that of random multifractal eigenfunctions, we compare Eqs.~(\ref{Crooks}) and (\ref{CrooksMF}) assuming that the drive protocol in (\ref{Crooks}) is time-reversal symmetric, therefore $\tilde P_w(W)=P_w(W)$. Using this comparison and the definition of $S$ and $\mathfrak{n}$ for multifractal wave functions we obtain
\begin{equation}\label{y-y}
y_{w}=\frac{(W-\Delta F)}{(k_{\rm B} T)\, \mathfrak{n}_{w}},\quad \Leftrightarrow \quad y=-\frac{\ln (N |\psi_{i}|^{2})}{\ln N},
\end{equation}
where the subscript $w$ stands for the distribution of work fluctuations. To determine the yet undefined parameter $\mathfrak{n}_{w}$ we use the following heuristic argument based on the above analogy. We note that  for a normalized eigenfunction on a lattice obeying $\sum_{i} |\psi_{i}|^{2}$=1 we have $|\psi_{i}|\leq 1$. This means that $y\geq -1$. A similar restriction for $y_{w}$  implies $(W-\Delta F)/k_{\rm B} T\geq -\mathfrak{n}_{w}$, that is
\begin{equation}\label{n_w}
\mathfrak{n}_{w}=E_{0}/k_{\rm B}T,
\end{equation}
where $(-E_{0})$ is the lower bound of the dissipated work. This result for the large-deviation parameter $\mathfrak{n}_{w}$ can be proven by a usual stochastic approach (see Eq.~(15) in the Supplementary Note 1) for the SEB governed by rate equations obeying detailed balance. However, we believe that it is valid generically for all  driven systems with the dissipated work bounded from below.
Thus the effective number (Eq.~(\ref{n_w})) of independent random variables in such driven systems  is inversely proportional to temperature $T$  and is easily controllable experimentally. This result is crucially important for experimental verification of our extension of the Jarzynski equality.
\subsection{Work generating function and extension of the Jarzynski equality}
\begin{figure*}[h!]
\centerline{%
\includegraphics[width=\textwidth]{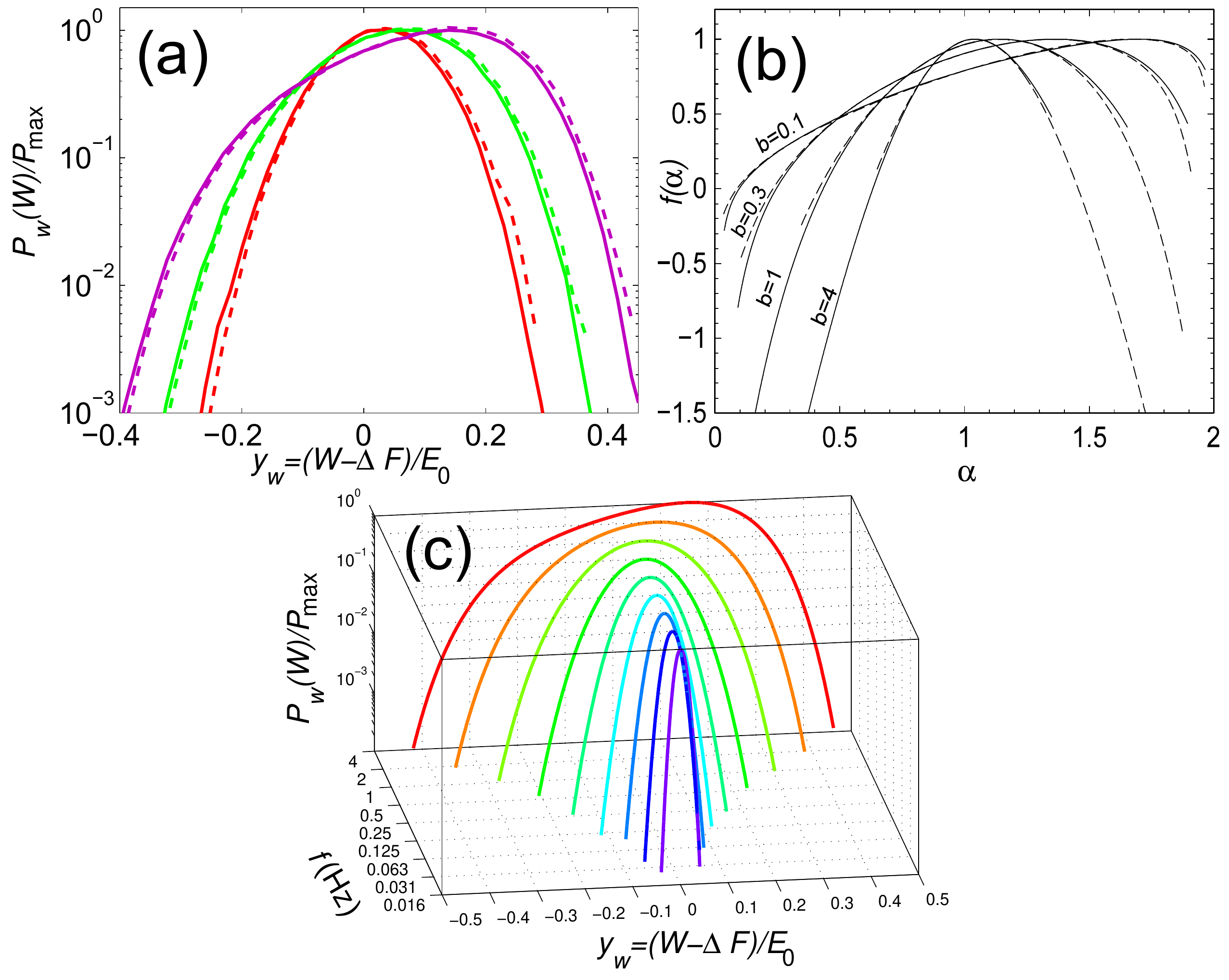}
}
\caption{{\bf Comparison of distributions of dissipated work and amplitudes of  random multifractal  wave functions.}
{\bf (a)} Distribution of the measured normalized dissipated work $(W-\Delta F)/E_0$ on the  logarithmic scale.
The width of the distributions increases with increasing drive frequency $f=1$ (red), $2$ (green), and $4$~Hz (violet)
at temperature $T=214$~mK.\\
 {\bf (b)} Multifractality spectrum $f(\alpha)$ of critical eigenfunctions in disordered systems close to the Anderson localization transition vs normalized logarithm of wave function intensity
$\alpha=-\ln|\psi_{i}|^2/\ln N$ for the power-law random banded matrix model with
the bandwidth $b = 0.1$, $0.3$, $1$, $4$ (adapted from \cite{MF_q_1-q_symmetry}).
This parameter is known to mimic the dimensionality of space in which the Anderson transition happens: $b\rightarrow 0$ corresponds to the limit of infinite dimensionality $d\rightarrow \infty$, or the Bethe lattice limit, while $b\rightarrow \infty$ corresponds to $d=2+\epsilon$, where $\epsilon\rightarrow +0$. In both panels (a) and (b), solid and dashed lines  correspond to $G(y)$, $G(-y)-y$ and $f(\alpha)$, $f(2-\alpha)+\alpha-1$, respectively, to demonstrate the symmetry~(\ref{MF-sym}).\\
{\bf (c)} Evolution of distribution of the normalized dissipated work $(W-\Delta F)/E_0$ on the logarithmic scale with decreasing drive frequency $f$ in a SEB with a superconducting external electrode for experimental system parameters and $T=214$~mK obtained theoretically from the rate equations (Eqs.~(3) and (4) in the Supplementary Note 1). The width of the distributions decreases with decreasing driving frequency $f$ (from red to violet curve).
The similar calculations for the SEB with the normal electrode give $b\sim \sqrt{k_{\rm B}T/(f\,\tau_{0}\,E_{0})}$, where $\tau_{0}$ is the characteristic relaxation time of the circuit.
Thus an effective bandwidth $b$ of the equivalent random matrix theory depends on the equivalent size of the matrix $N={\rm exp}(E_{0}/k_{\rm B}T)$.
While the limit $T\rightarrow 0$ always corresponds to the limit $b\rightarrow 0$, the limit $f\rightarrow 0$ at a fixed $T$ corresponds to the adiabatic limit $b\rightarrow \infty$.
}
\label{Fig:PDF-analogy}
\end{figure*}

\begin{figure*}[h!]
\centerline{
\includegraphics[width=\textwidth]{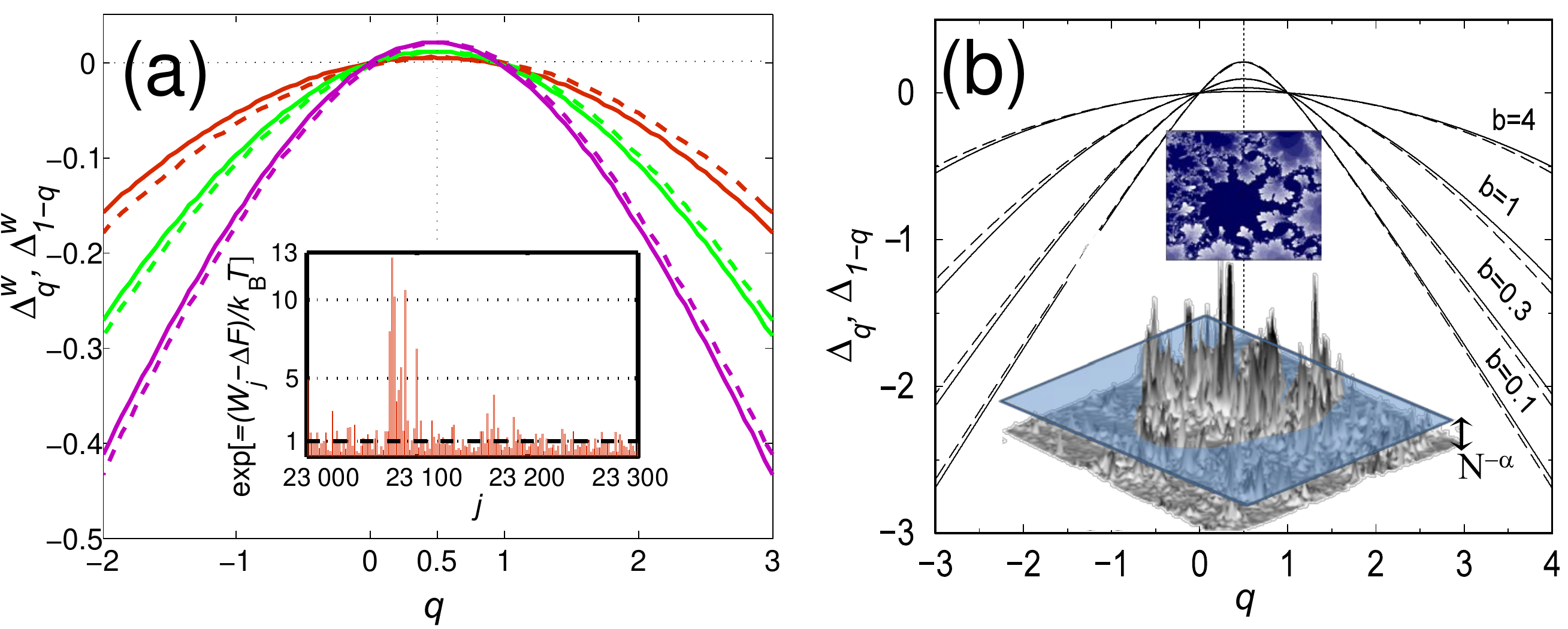}
}
\caption{{\bf Comparison of $\Delta^{w}_{q}$ for dissipated work and multifractal critical exponents $\Delta_{q}$.}\\
{\bf (a)} The measured function $\Delta_q^{w}(T)$ in Eq.~(\ref{generalization}) at
drive frequencies $f=1$ (red), $2$ (green), and $4$~Hz (violet) and\ temperature $T=214$~mK.
(inset) A plot of the exponent of the dissipative work $e^{-(W-\Delta F)/k_{\rm B}T}$ vs the drive realization number $j$ at a drive frequency $f=1$~Hz and temperature $T=214$~mK.
In most of the drives the exponent is smaller than 1, which corresponds to $W>\Delta F$, as the Second Law requires for averages. However, there are rare events seen as high spikes when $\Delta F -W> k_{\rm B} T $.\\
{\bf (b)} Multifractal exponents $\Delta_q$ for the same model and parameters as in Fig.~\ref{Fig:PDF-analogy}~(b) (adapted from \cite{MF_q_1-q_symmetry}).
In both panels a small difference between $\Delta_q$ (solid lines)
and $\Delta_{1-q}$ (dashed lines) violating the symmetry~(\ref{sym-delta}) is due to experimental (in panel (a)) or numerical (in panel (b)) errors.
(bottom inset) A plot of a typical amplitude $|\psi_i|^{2}$ of the critical wave function in a 2D lattice with $N$ sites cut at a certain level $|\psi_i|^{2}=N^{-\alpha}$ (adapted from \cite{WF_Yakubo}).
(top inset) The map of the region in space where $|\psi|^{2}>N^{-\alpha}$ is a fractal of the Haussdorf dimension $d_{\rm h}(\alpha)=2\,f(\alpha)<2$. Multifractality implies a dependence of $d_{\rm h}$ on $\alpha$, or on the cutoff level $N^{-\alpha}$ (adapted from \cite{2007_Cuevas_Kravtsov}).
}
\label{Fig:Delta-analogy}
\end{figure*}

With the established  physical meaning of $\mathfrak{n}_{w}$, the analogy between the work distribution in driven systems and the multifractal statistics of random eigenfunctions becomes complete. It is illustrated in Figs.~\ref{Fig:PDF-analogy} and \ref{Fig:Delta-analogy}.

A remarkable property of the LDA (\ref{LDA}) is that the average of  $\langle e^{-q\, S}\rangle \sim e^{-\mathfrak{n}\,\tilde{\Delta}_{q}}$ is an exponential function of $\mathfrak{n}\gg 1$, where  $\tilde{\Delta}_{q}={\rm min}_{y}\, \{q\,y + G(y)\}$.\cite{Touchette,Esposito_PRE88} Given that $\mathfrak{n}=\ln N$ this implies a power-law scaling with $N$ of the moments $\langle N^{q} |\psi_{i}|^{2q}\rangle\sim N^{-\Delta_{q}}$ (with $\Delta_{q}\Leftrightarrow\tilde{\Delta}_{q}$) of  random wave functions near the critical point of the Anderson localization transition. When applied to the statistics of work, the exponential dependence on $\mathfrak{n}_{w}=E_{0}/k_{\rm B} T$ results in the following relation for the work generating function $F(q)= \langle e^{-q\,(W-\Delta F)/k_{\rm B} T}\rangle $:
\begin{widetext}
\begin{equation}\label{generalization}
\ln \langle e^{-q\,(W-\Delta F)/k_{\rm B} T}\rangle \equiv -(E_{0}/k_{\rm B}T)\, \Delta^{w}_{q}(T)\; \stackrel{T\to 0}{\longrightarrow}\; -(E_{0}/k_{\rm B}T)\, \Delta^{w}_{q} \ ,
\end{equation}
\end{widetext}
where the limit $\Delta^{w}_{q}$ is independent of temperature.
Equation (\ref{generalization}) is the main theoretical result of our work, where we claim that the logarithm of the work generating function is linear in $E_{0}/k_{\rm B} T\gg 1$ with $\Delta^{w}_{q}$ being a non-trivial function of a real $q$. It generalizes the Jarzynski equality which corresponds to $q=1$ and $\Delta^{w}_{q=1}(T)=0$. Apparently, we have also $\ln \langle 1 \rangle \propto \Delta^{w}_{q=0}(T)=0$. One can easily show using Eq.~(\ref{MF-sym}) and the definition of $\Delta^{w}_{q}$ that the symmetry
\begin{equation}\label{sym-delta}
\Delta^{w}_{q}(T) = \Delta^{w}_{1-q}(T)
\end{equation}
holds both for $\Delta^{w}_{q}$ and for $\Delta^{w}_{q}(T)$.
This symmetry has its counterpart for the critical exponents $\Delta_{q}$ that determine the scaling with $N$ of the moments of random critical wave functions.
The limit $\Delta^{w}_{q}$ of $\Delta^{w}_{q}(T)$ at $E_{0}/k_{\rm B} T\rightarrow\infty$
(at a fixed drive frequency $f$)
is expected to be robust to changing the details of the driven system and the drive protocol.
For a driven two-level system, described by rate equations (Eqs.~(3, 4) in the Supplementary Note 1) and obeying detailed balance  $\Gamma_{+}(U)=\Gamma_{+}(-U)\,e^{U/k_{\rm B} T}$ for the up (down) transition rates $\Gamma_{+}$ ($\Gamma_{-}$), standard stochastic dynamics calculus confirms the main result Eqs.~(\ref{n_w}, \ref{generalization}), with $\Delta^{w}_{q}$ having always the same asymptotic behavior $\Delta^{w}_{q}\approx \frac{1}{2}-\left|q-\frac{1}{2} \right|$ at large enough $|q|\gg q_{c}$ (see Eq.~(18) in the Supplementary Note 1). This form of $\Delta^{w}_{q}$ corresponds to the limit of infinite dimensions, or the Bethe lattice limit \cite{Kravtsov_Altshuler_N-expansion}, in the problem of the random critical wave functions.
Note that the universal behavior of $\Delta^{w}_{q}$ (and the corresponding behavior of $P_w(W)$) is reached only in the limit $T\rightarrow 0$, with all other parameters of the system and drive being fixed. If, however, the temperature is low but fixed then there always exists a sufficiently low drive frequency $f$ such that the dissipated work distribution tends to a $\delta$-function, as the adiabatic limit requires.\cite{SEB_heat_stats} For a single-electron box with a superconducting electrode the range of such frequencies could be extremely low at temperatures $k_{\rm B} T\ll \Delta_{\rm S}$ with $\Delta_{\rm S}$ being a superconducting gap in the island (see Fig.~\ref{Fig:PDF-analogy}(c)).
\subsection{Experimental verification for a single-electron box}
\begin{figure*}[h!]
\centerline{
\includegraphics[width=\linewidth]{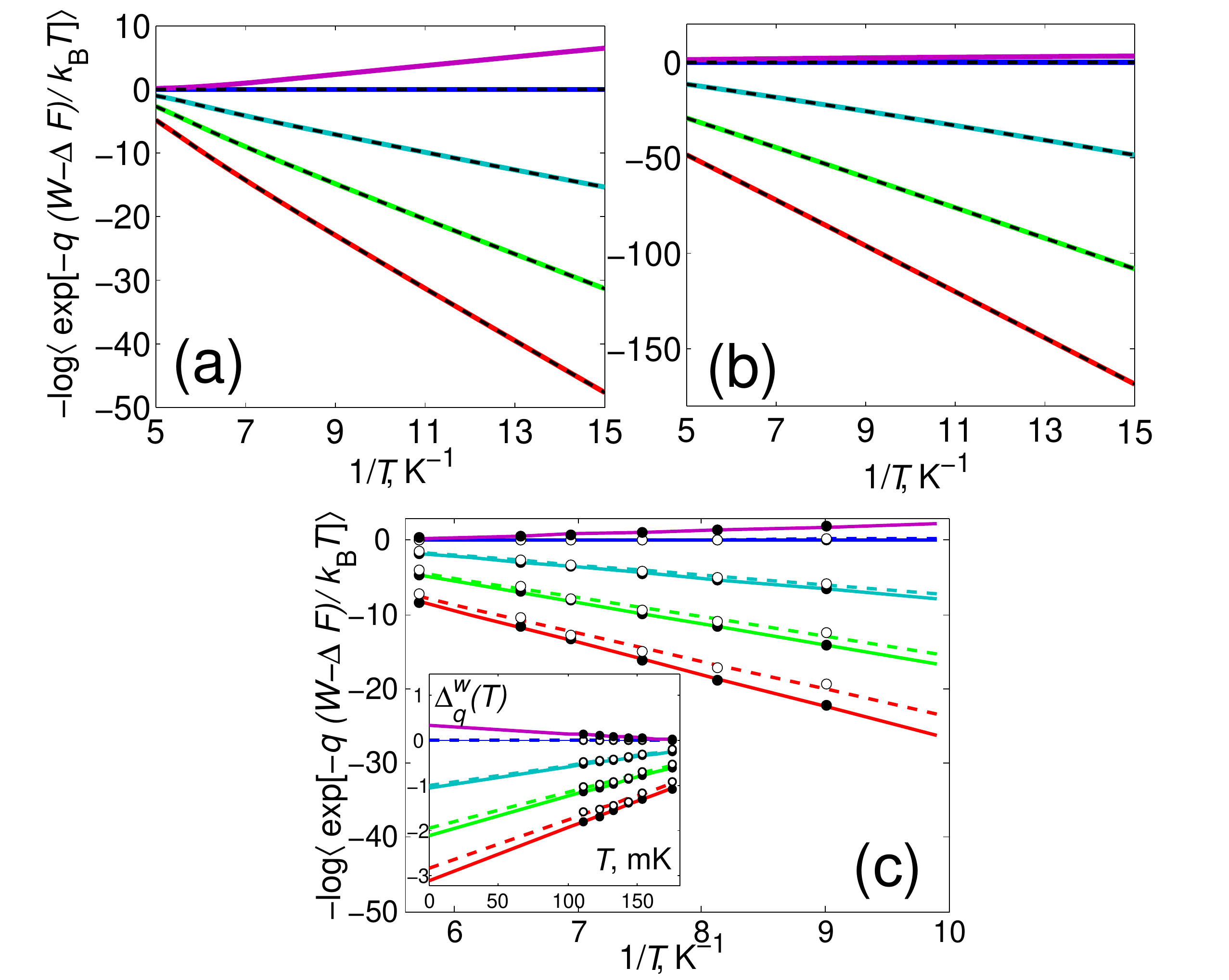}
}
\caption{{\bf The dependence in $T^{-1}$ of the logarithm of the work generating function and its symmetry in the moment order $q$.} In panels (a) and (b) the theoretical $T^{-1}$~--~dependence obtained from the rate equations for a SEB with (a) a superconducting and (b) a normal external electrode is shown. Panel (c) demonstrates the experimental test of this dependence. In all panels the dependencies become linear at large values of $T^{-1}$.
The dashed (solid) lines correspond to the pairs of moments $\{q , 1-q\}$ related by symmetry.
The curves from bottom to top correspond to $\{ 4,-3\}$, $\{3,-2\}$, $\{2,-1\}$, $\{ 1,0\}$ and $q=\frac{1}{2}$.
In panel (c) the dashed (solid) lines of the same color are linear fits of the experimental data shown by open (solid) circles corresponding to $q\quad (1-q)$.
(inset) The linear in $T$ extrapolation for the function $\Delta_q^w(T)$. The notations are the same as in panel (c).
}
\label{Fig:linearity_1_T}
\end{figure*}

\begin{figure*}[h!]
\centerline{
\includegraphics[width=\linewidth]{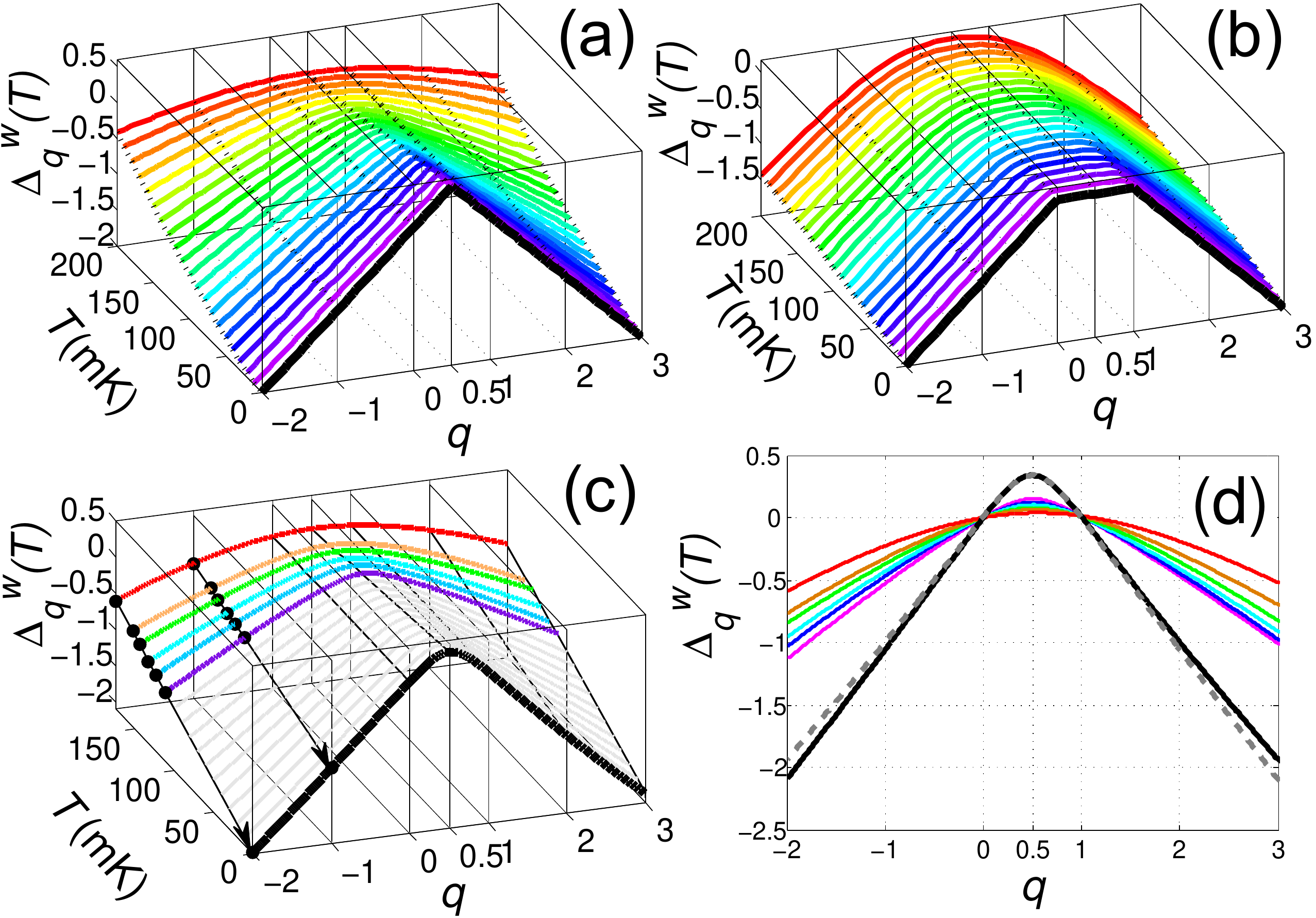}
}
\caption{{\bf Evolution of $\Delta^{w}_{q}(T)$ with decreasing temperature $T$ in a single-electron box.} Panels (a) and (b) show the theoretical evolution for a SEB with (a) a superconducting external electrode for $\Delta_{\rm S}=E_{\rm C}$  and (b) a normal external electrode obtained from the rate equations. The limiting $\Delta^{w}_{q}$ is of (a) triangular, (b) trapezoidal form.
Experimental data is shown in panels (c) and (d). (c) Experimentally obtained $\Delta_q^w(T)$ for temperatures $T=111$, $123$, $133$, $144$, $153$, and $175$~mK are plotted as functions of  $q$ (colored solid lines).
The solid black curve $\Delta^{w}_{q}$ is obtained by linear in $T$ extrapolation of the experimental data to zero temperature. The thin gray curves and dotted arrows demonstrate the linear extrapolation.
Panel (d) shows the verification of the symmetry (\ref{sym-delta}). Here the dashed gray curve shows $\Delta_{1-q}^{w}$, the other notations are the same as in panel (c).
}
\label{Fig:Delta-evolution}
\end{figure*}

The general theory above can be applied to a driven SEB, which is a small metallic island connected to an external electrode with a tunnel junction.
The free electrons on the SEB island and the electrode form a particle bath, assumed to be at thermal equilibrium at temperature  $T$.\cite{SEB_JE_test_Saira,SEB_Jukka}
A standard rate equation approach \cite{Averin_Likharev86,Likharev87}
which is essentially classical and based on the picture of sequential tunneling of electrons,
confirms our main result (\ref{generalization}) and the symmetry (\ref{sym-delta}) (see Supplementary Notes 1 and 2). This theory gives a linear in $T^{-1}$ low-temperature behavior of the cumulant generating function [left-hand side of Eq.~(\ref{generalization})], as shown in Fig.~\ref{Fig:linearity_1_T}(a, b).
We consider two different cases as examples belonging to different universality classes
marked by drastically different dependence of the 
tunneling rate $\Gamma_{+}(U)$  on the drive voltage $U\gg k_{\rm B} T$:
a SEB with normal island  and (a) a superconducting external electrode ($\Gamma_{+}\sim e^{-\Delta_{\rm S}/k_{\rm B} T}\sqrt{\Delta_{\rm S} k_{\rm B} T}\,(1+e^{U/k_{\rm B} T})$, $\Delta_{\rm S}>U$ ) or (b) a normal external electrode ($\Gamma_{+}\sim U\,\sinh^{-1}(U/2k_{\rm B} T)\,e^{U/2k_{\rm B} T}$).
The evolution of $\Delta^{w}_{q}(T)$ with temperature in both cases is shown in Fig.~\ref{Fig:Delta-evolution}. 
The limiting $\Delta^{w}_{q}$ appears to be of triangular shape in case (a), and of trapezoidal shape in case (b).

The main quantum effects which are beyond the rate equation approach, are the elastic co-tunneling\cite{Averin_Nazarov_Cotunneling_1990} and the Andreev tunneling\cite{Averin_Pekola_PRL_2008_Andreev_tun}. Estimations show (see Eqs.~(3) the Supplementary Note 3) that for SEB at our experimental conditions they may become relevant at low temperatures $T<T^{*}\sim 60$~mK. We believe that these effects merely renormalize the parameter $E_{0}$ and the function $\Delta^{w}_{q}$   and do not change the $1/T$ behavior in Eq.~(\ref{generalization}). Further investigations are necessary to check the validity of this conjecture.

For an experimental verification of our main result (\ref{generalization}) and the symmetry (\ref{sym-delta}) we use
a SEB formed by two metallic islands, of which one is normal and the other one is superconducting with energy gap $\Delta_{\rm S}$. As a two-island SEB is only capacitively coupled to the environment, it is less influenced by external noise from higher temperature stages of the setup. Otherwise its behavior is identical to a normal one-island SEB with a superconducting ``external electrode''. The measured structure is described in Refs.~\cite{SEB_JE_test_Saira,Koski_MD-1,Koski_MD-2}.
We used aluminum and copper as a superconductor and a normal metal, respectively, and apply magnetic field to increase the tunneling rates through the junction by suppressing the gap $\Delta_{\rm S}$, see the Supplementary Note 4 for details.
The Hamiltonian $H(n,n_{\rm g})=E_{\rm C}\,(n^{2}-2n\, n_{\rm g})$ of the SEB consists of the charging energy of the island with an integer number of excess electrons $n$ and the interaction with the source of the gate voltage
$V_{\rm g}$ controlling the gate charge $n_{\rm g} = C_{\rm g} V_{\rm g}/e$ through the capacitance $C_{\rm g}$.
The energy required to charge the island with a single electron $-e$ is  $E_{\rm C} = e^2 / 2 C_\Sigma$, where $C_\Sigma$ is the total capacitance of the island. In this experiment, we apply a sinusoidal modulation $n_{\rm g}(t) = \frac{1}{2} - \frac{1}{2}\cos(2\pi f t)$ and consider a monotonous segment of $n_{\rm g}(t)$ from $0$ to $1$ as a single realization of the process $0<t<(2f)^{-1}$. We focus on the large Coulomb energy limit $E_{\rm C} \gg k_{\rm B} T$, in which $n$ is restricted to two values, $n=0$ and $n=1$. In this case, the dissipated work is determined from the trajectory of $n(t)$ by \cite{SEB_heat_stats,SEB_work_stats}
\begin{equation}\label{work-def}
W - \Delta F =-E_{\rm C}\int_{0}^{1}(2n-1)\,dn_{\rm g} .
\end{equation}
Like in the textbook example of a moving piston where the volume $V(t)$ of the gas is controlled deterministically and the pressure $p(t)$ experiences fluctuations due to collisions of gas atoms with the piston, the gate voltage $n_{\rm g}(t)$ is a deterministic function whereas $n(t)$ experiences telegraph fluctuations. These fluctuations are detected by a nearby charge-sensitive single electron transistor.
The dissipated work is computed from Eq.~(\ref{work-def}) and its statistics over repeated identical driving protocols is studied. Here the lower bound $-E_{0}$ of the dissipated work is determined by the Coulomb energy $E_0=E_{\rm C}$.

The experimental PDFs of work for few different drive frequencies are presented in Fig.~\ref{Fig:PDF-analogy}. This plot demonstrates the dependence of the PDF on the frequency which is reminiscent of the dependence on the bandwidth $b$ of the corresponding PDF for random multifractal wave functions for the power-law banded random matrices.\cite{Mirlin-review} Using this PDF one can compute the $q$th moments of $e^{-(W-\Delta F)/k_{\rm B} T}$ for different values of the parameter $q$  and find the function $\Delta^{w}_{q}(T)$ from Eq.~(\ref{generalization}) (see Fig.~\ref{Fig:Delta-analogy}). In both figures the charging energy of the SEB determining the dissipated work (\ref{work-def}) is $E_{\rm C} = 167\pm4$~$\mu$eV, while the bath temperature is $T = 214$~mK. The drive
frequencies are indicated in the figures.

Next we check experimentally the linear in $E_{0}/k_{\rm B} T$ low-temperature dependence in Eq.~(\ref{generalization}) and the symmetry of Eq.~(\ref{sym-delta}). The results are presented in Fig.~\ref{Fig:linearity_1_T}(c). The corresponding theoretical curves are given in Fig.~\ref{Fig:linearity_1_T}(a). Note a good linearity of experimental data for the negative $q$ (full circles, solid lines) and a much larger scatter of it (open circles) for the large positive $q$ which corresponds to rare events with $W-\Delta F<0$.
The linear in $T$ evolution of $\Delta^{w}_{q}(T)=\Delta^{w}_{q}+c_{q}\,(k_{\rm B} T/E_{0})$ is demonstrated experimentally  in Fig.~\ref{Fig:Delta-evolution}(c). Its counterpart for the random eigenfunction problem is the evolution with the system size $N$ linear in $1/\ln N$ which was used recently in Ref. \cite{Kravtsov_Altshuler_N-expansion} to find the spectrum of fractal dimensions $f(\alpha)$ extrapolated to the infinite system size.  Similarly to this work, the limiting function $\Delta^{w}_{q}$ is obtained by the linear in $T$ extrapolation to $T\rightarrow 0$ (see the inset in Fig.~\ref{Fig:linearity_1_T}(c)).
In both figures the charging energy is $E_{\rm C} = 111\pm4$~$\mu$eV, the drive frequency is $f= 4$~Hz, while the temperatures are indicated in the figures.
The estimated \cite{1964_Skalsky,1965_Maki,2003_DOS_in_H} value of the superconducting energy gap $\Delta_{\rm S} = 96\pm 11$~$\mu$eV in applied magnetic field is rather close to $E_{\rm C}$ in this case. The corresponding extrapolated function  $\Delta^{w}_{q}$ shown in Fig.~\ref{Fig:Delta-evolution}(c)  is close to the triangular form obtained theoretically from the rate equations in the ideal case $\Delta_{\rm S}=E_{\rm C}$ and shown in Fig.~\ref{Fig:Delta-evolution}(a), albeit it is somewhat rounded
on the top following a trend towards the trapezoidal form shown in Fig.~\ref{Fig:Delta-evolution}(b).
The asymptotic behavior of the extrapolated function $\Delta_q^w$ at $q>1$ or $q<0$ is close to the theoretically predicted asymptotics $\Delta_q^w = 1/2-|q-1/2|$, linear in $q$ with unit slope, supporting the linear in $T$ extrapolation.

\section{Discussion}
In conclusion, we have shown that the analogy between the statistics of random critical wave functions and that of the work dissipated in driven systems, is very suggestive. Its predictions are fully confirmed theoretically and experimentally by studying stochastic dynamics in a driven SEB described by rate equations obeying detailed balance. Thus one of the most difficult problems in quantum mechanics of disordered systems   turns out to be analogous to one of the simplest problem in classical stochastic Markovian dynamics.
In particular, the physical origin of the symmetry (\ref{sym-delta}) is somewhat unclear 
in the problem of Anderson localization (see though \cite{Fyodorov-Savin}). At the same time, the corresponding symmetry for driven systems is a consequence of the Crooks relation or, equivalently, of detailed balance for rate equations. This might suggest that there would be a stochastic description for the critical random eigenfunction problem by an equivalent Markovian process with detailed balance.
One can see a remote analogy of  such correspondence in the Schramm-Loewner evolution which maps fractal phase boundaries in 2D critical systems onto a simple random walk on a line \cite{Schramm,SLE-rev}.







\vspace{1cm}

\acknowledgements
We thank Prof.~F.~Hekking, Prof. B.~L.~Altshuler for useful discussions and Prof.~A.~S.~Mel'nikov for advice in improving the manuscript.
This work has been supported in part by
the Academy of Finland (projects no. 250280 and 272218),
the European Union Seventh Framework Programme INFERNOS (FP7/2007-2013) under grant agreement no. 308850,
the Russian president foundation (project no. SP-1491.2012.5),
and the V{\"a}is{\"a}l{\"a} Foundation. V.E.K. acknowledges the hospitality of LPTMS of University of Paris Sud at Orsay and support under the CNRS grant ANR-11-IDEX-0003-02 Labex PALM, project MultiScreenGlass.
We acknowledge the availability of the facilities and
technical support by Otaniemi research infrastructure for
Micro and Nanotechnologies (OtaNano).
\subsection{Author Contributions}
J.~V.~K., O~-~P.~S., and J.~P.~P. conceived and designed the experiments;
J.~V.~K. and O~-~P.~S. performed the experiments;
I.~M.~K., J.~V.~K., and O~-~P.~S. analyzed the data.
I.~M.~K., V.~E.~K., and J.~P.~P. contributed with materials/analysis tools;
I.~M.~K., J.~V.~K., V.~E.~K., and J.~P.~P. wrote the paper.
\subsection{Additional Information}
Supplementary notes are available in the online version of the paper.
Reprints and permissions information is available at www.nature.com/reprints.
Correspondence and requests for materials should be addressed to I.~M.~K.~(email: ivan.khaymovich@aalto.fi).
\subsection{Competing Financial Interests}
 The authors declare that they have no
competing financial interests.

\onecolumngrid

\section*{\Large Supplementary Information}

\section{Supplementary Note 1 Statistics of work for a single-electron box}\label{SM_sec:work-stats}
In this part we give a sketch of the calculation of PDF $P_w(W_{\rm d})$ of dissipated work $W_{\rm d}=W-\Delta F$ in a SEB
which confirms Eq.~(7) 
from the main text and the asymptotics $\Delta_q^w = 1/2-|q-1/2|$ for a simple Markovian system described by rate equations.
For the sake of clarity we consider the time-reversal (anti)symmetric protocol of the gate voltage $n_{\rm g}(t)$ monotonously increasing from $0$ to $1$ in time $0<t<\tau=(2f)^{-1}$, i.e., $n_{\rm g}(\tau-t)=1-n_{\rm g}(t)$, where the symmetry (1) 
 is satisfied with $\tilde P_w(W)=P_w(W)$. As was mentioned in the main text we also focus on the large Coulomb energy limit $E_{\rm C} = e^2/[2(C_{\rm g}+C)]\gg k_{\rm B} T$. Then the excess number of electrons $n$ on the island is restricted to two values $n=0$ and $n=1$ in this range of gate voltage, $0\leq n_{\rm g}(t)\leq 1$.

The Hamiltonian $H(n,n_{\rm g}) = E_{\rm C}(n^2-2 n_{\rm g} n)$ of a SEB mentioned in the main text determines the dissipated work as follows \cite{SM:SEB_work_stats}:
\begin{equation}\label{SM:W_diss}
W_{\rm d}[n(t),n_{\rm g}(t)]=W-\Delta F = -E_{\rm C} \int_0^1 (2n-1) dn_{\rm g} \ ,
\end{equation}
with the (thermodynamical) work $W = \int \frac{\partial H}{\partial n_{\rm g}} dn_{\rm g}$, and the free energy difference $\Delta F = F(1)-F(0)$, where $\beta
F(n_{\rm g}) =-\ln\left[\sum_n e^{-\beta H(n,n_{\rm g})}\right]$ and $\beta = (k_{\rm B} T)^{-1}$. For the chosen  drive protocol the minimal dissipated work $-E_0=\min W_{\rm d} = -E_{\rm C}$
is negative with the absolute value coinciding with the Coulomb energy.

During the ramp of $n_{\rm g}(t)$ different stochastic trajectories $n(t)$ of the charge state occur with alternating jumps of $n$ either from $0$ to $1$ or vice versa. Each trajectory unambiguously determines the dissipated work for a realization, see Eq.~(\ref{SM:W_diss}).
To find the distribution of dissipated work
\begin{eqnarray}\label{SM:PDF-traject}
P_w(W_{\rm d}) = \sum_{k=0}^{\infty}\sum_{\eta=0}^1 \int P_k(\eta,t_1,\ldots, t_k) 
\delta\left(W_{\rm d}[n(t),n_{\rm g}(t)]-W_{\rm d}\right)dt_1\cdot\ldots\cdot dt_k
\end{eqnarray}
one calculates the probability $P_k(\eta,t_1,\ldots, t_k)$ of realizing a trajectory $n(t)$ starting at $n(0)=\eta=\overline{0,1}$ which has $k$
jumps occuring at time instants $t_1,\ldots, t_k$.
For this purpose we solve the master equation for the occupation probabilities $p_n$ of the two charge states
\begin{equation}\label{SM:rate_eq}
\frac{dp_1}{dt} = \Gamma_+(t) p_0 - \Gamma_-(t) p_1, \quad p_0 = 1-p_1 \ ,
\end{equation}
with the equilibrium initial state $p_1(0) = \Gamma_+(0)/[\Gamma_+(0)+\Gamma_-(0)]\approx e^{-\beta E_{\rm C}}$.
Here the rates $\Gamma_\pm(t)$ of electron tunneling into/out of the island can be written as follows $\Gamma_\pm(t)=\Gamma[\pm U(t)]$ with the function
\begin{equation}\label{SM:Gamma[U]}
\Gamma[U] =\frac{1}{e^2 R_{\rm T}}\int\limits_{-\infty}^{\infty}\nu(E) f_T(E) [1-f_T(E+U)]dE \ ,
\end{equation}
monotonically increasing with  $U(t) = H(0,n_{\rm g})-H(1,n_{\rm g})$:
\begin{equation}
U(t) = E_{\rm C} [2 n_{\rm g}(t)-1].
\end{equation}
Here $R_{\rm T}$ is the tunnel resistance of the contact and we consider a normal-metal island with the constant density of states, while
the external electrode can be either in a normal (N) or in a superconducting (S) state with the normalized density of states equal to $\nu(E)=1$ or $\nu(E)=\left|{\rm Re}
\frac{E}{\sqrt{E^2-\Delta^2}}\right|$, respectively.
We also assume that electrons in the island and in the electrode thermalize quickly enough to have the same Fermi distributions of energy   $f_T(E)=\left(e^{\beta E}+1\right)^{-1}$ with the temperature $T$ of the single bath.
One of the main consequences of the latter assumption about thermalization and Eq.~(\ref{SM:Gamma[U]}) is the detailed balance of the tunneling rates
\begin{equation}\label{SM:detail_balance}
\Gamma_+(t)/\Gamma_-(t) = e^{\beta U(t)} \ ,
\end{equation}
which eventually results in the Crooks relation (1) 
 for dissipated work (\ref{SM:W_diss}) in the system described by rate equation
\cite{SM:Crooks_1998,SM:Monthus_Berche_MF_analogy} and driven by time-reversal (anti)symmetric drive
\begin{equation}\label{SM:drive_symmetry}
U(\tau-t)=-U(t), \; \texttt{when} \quad \Gamma_+(\tau-t)=\Gamma_-(t) \ .
\end{equation}

In the low temperature limit $k_{\rm B} T/E_{\rm C}\rightarrow 0$ (at a fixed drive frequency $f=(2\tau)^{-1}$) we can restrict our consideration to zero- and one-jump trajectories starting from the ground state $n(0)=0$.
The probabilities of considered trajectories can be calculated in a similar way as in Ref.~\onlinecite{SM:SEB_heat_stats}
\begin{eqnarray}
P_0(\eta) &=& p_{\eta}(0)e^{-A} \ , \\
P_1(0,t_1) &=& p_{0}(0)e^{-I(t_1)}\Gamma_{+}(t_1)  \ ,
\end{eqnarray}
where $A=\int_0^\tau \Gamma_+(t) dt$ and $I(t_1)=\int_0^{t_1} \Gamma_{+}(t)dt+\int_{t_1}^\tau \Gamma_{-}(t)dt$ is symmetric, $I(\tau-t)=I(t)$, due to (\ref{SM:drive_symmetry}), it is bounded $0\leq I\leq A$ and $I(0)=A$.
The validity of considering only zero- and one-jump trajectories can be verified using the following estimate for the total probability of  trajectories with $k>1$ jumps,
\begin{equation}
P_{k>1} = 1-\sum_{\eta=\pm}P_0(\eta)-\int\limits_0^\tau P_1(0,t_1)dt_1 \leq I(\tau/2) \ ,
\end{equation}
which vanishes in the limit $T\to 0$.
Indeed,
\begin{eqnarray}
\int\limits_0^\tau P_1(0,t_1)dt_1&\approx&\int\limits_0^\tau e^{-I(t_1)}\Gamma_{+}(t_1) dt_1 =\nonumber \\
\int\limits_{U(t)>0} e^{-I(t_1)}\left[\Gamma_{+}(t_1)+\Gamma_{-}(t_1)\right] dt_1 &\geq&
\int\limits_{U(t)>0} e^{-I(t_1)}\dot{I}(t_1) dt_1 = e^{-I(\tau/2)}-e^{-A} \ ,
\end{eqnarray}
and
\begin{equation}
P_{k>1} \leq 1-e^{-I(\tau/2)}\leq I(\tau/2) \ ,
\end{equation}
where
\begin{equation}
I(\tau/2) = 2\int\limits_{U(t)>0}\Gamma[U(t)] e^{-\beta U(t)}dt\lesssim \frac{\Gamma_0 \tau}{\beta E_{\rm C} n_{\rm g}'(\tau/2)}\to 0 \
\end{equation}
and $n_{\rm g}'(\tau/2) = \partial n_{\rm g}/\partial(t/\tau)$ is the derivative of the gate voltage $n_{\rm g}$ over the normalized time.
Here we used the symmetry~(\ref{SM:drive_symmetry}) and the natural assumption that the maximal value of the tunneling rates $\Gamma[E_{\rm C}]$ is bounded to $\Gamma[E_{\rm C}]<\Gamma_0=\rm const$ in the considered limit of $k_{\rm B} T/E_{\rm C}\rightarrow 0$.

As a result, Eq.~(\ref{SM:PDF-traject}) yields
\begin{eqnarray}\label{SM:PDF1}
 P_w(y_w)\approx e^{-A}\left[\delta(y_w-1)+e^{-\beta E_{\rm C}}\delta(y_w+1)\right]+
 \frac{\tau}{2}\Gamma[E_{\rm C}\cdot y_w] e^{-I(y_w)} \ ,
\end{eqnarray}
where $y_{w}$ is defined in Eq.~(5) of the main text.
The singular part of $P_w$ corresponds to the trivial jumpless trajectories. They make a contribution to $\Delta_q^w(T)-\Delta_q^w(0)=O(1/(\beta E_0))$ in Eq.~(7) 
from the main text which is subleading.
The regular part of $P_{w}(y_{w})$ can be compared with the large deviation ansatz of Eq.~(2) 
from the main text by taking the limit $G(y)=\lim _{\mathfrak{n}\rightarrow\infty} G(y,\mathfrak{n})$ of $G_{w}(y_{w}, \mathfrak{n}_w)\equiv -\ln [P_{w}(y_{w})]/\mathfrak{n}_w$ as
\begin{equation}\label{SM:f_w(1+y_w)}
G_w(y_w,\mathfrak{n}_w)=  -\frac{\ln \gamma(y_w)}{\mathfrak{n}_w}+\frac{I(y_w)-\ln (\Gamma[E_{\rm C}]\tau/2)}{\mathfrak{n}_w} \ .
\end{equation}
Here $\mathfrak{n}_w=\beta E_{\rm C}$, $\gamma(y_w)=\Gamma[E_{\rm C} y_w]/\Gamma[E_{\rm C}]$ and the second fraction vanishes when $T\to 0$.
The first term gives the main contribution to $G_w(y_w)$ and for the normal external electrode  $\gamma(y_w)={y_w}(1-e^{-\beta E_{\rm C} y_w})^{-1}$  we obtain
\begin{equation}
G_w(y_w)=\left\{-y_w,\; -1\leq y_w\leq 0\atop \;\;\;\;0,\;\;\;\;0\leq y_w \leq 1\right. \ ,
\end{equation}
while for the superconducting external electrode with $\gamma(y_w)=e^{-\beta E_{\rm C}}(1+e^{\beta E_{\rm C} y_w})$ the function $G_w(y_w)$ takes the form
\begin{equation}
G_w(y_w)=\left\{\;\;\;\;\;\;\;1,\; -1\leq y_w\leq 0\atop 1-y_w,\;\;\;\;0\leq y_w \leq 1\right. \ .
\end{equation}

The corresponding limiting  $\Delta_q^w = \min_{y_w}\, \{y_w\,q + G(y_w)\}$ is
\begin{equation}\label{SM:Delta-NIS}
\Delta_q^w \to \frac{1}{2}-\left|q-\frac{1}{2}\right| \ ,
\end{equation}
for the case of the superconducting external lead, while for the normal external lead the positive part of $\Delta_{q}^{w}$ in Eq.~(\ref{SM:Delta-NIS}) at $0<q<1$ is replaced by 0. In both cases the asymptotic behavior $\Delta^{w}_{q}\approx \frac{1}{2}-\left|q-\frac{1}{2} \right|$ for $q<0$ or $q>1$ holds true.

In general for low finite temperatures, $k_{\rm B} T\ll E_{\rm C}$, the averaging in Eq.~(7) 
from the main text can be calculated using the saddle-point approximation and we obtain
\begin{equation}\label{SM:Delta_q}
\Delta_q^w(T) = \min_{y_w}\, \{y_w\,q + G(y_w,\mathfrak{n}_w)\}=\Delta_q^w+c_q(T)/\mathfrak{n}_w \ .
\end{equation}
Here the last term $c_q(T)/\mathfrak{n}_w$ originates from the second fraction in Eq.~(\ref{SM:f_w(1+y_w)}) with bounded $c_q(T)<c_q^{\max}=\rm const$ and therefore it is a subleading term.
As a result, rewriting Eq.~(7) 
from the main text in the following form
\begin{eqnarray}
\ln \langle e^{-q\,(W-\Delta F)/k_{\rm B} T}\rangle = -(E_{0}/k_{\rm B}T)\, \Delta^{w}_{q} + c_q(T) 
= -(E_{0}/k_{\rm B}T)\, \Delta^{w}_{q} + O(1) \ ,\label{SM:generalization}
\end{eqnarray}
we prove the linear behavior of the l.h.s. in $E_{0}/k_{\rm B}T$.

The linear $T$-expansion of $\Delta_q^w(T)$, Eq.~(\ref{SM:Delta_q}), to $\Delta_q^w$ is possible in the range of temperatures, where $c_q(T)$ is close to its limiting value $c_q(0)$, i.e., when $c_q(T)\approx c_q(0)$.

\section{Supplementary Note 2 Derivation of Eq.~(8) 
}\label{SM_sec:sym-delta}
The symmetry  $\Delta_q^w(T) = \Delta_{1-q}^w(T)$ of Eq.~(8) 
 can be proved simply using Crooks relation for time-reversal symmetric drive protocol (4): 
  $P_w(y_w) = e^{\mathfrak{n}_w y_w}P_w(-y_w)$. Indeed,
\begin{eqnarray}
e^{-\mathfrak{n}_w \Delta_q^w(T)} = \int e^{-q \mathfrak{n}_w y_w}P_w(y_w) dy_w = 
\int e^{(1-q) \mathfrak{n}_w y_w}P_w(-y_w) dy_w =
\int e^{-(1-q) \mathfrak{n}_w y_w'}P_w(y_w') dy_w' = e^{-\mathfrak{n}_w \Delta_{1-q}^w(T)}\label{SM:sym-delta} \ .
\end{eqnarray}

\section{Supplementary Note 3 Estimation of the coherent processes in SIN SEB}\label{SM_sec:higher_order_processes}
In this section we estimate temperature range where coherent phenomena, namely, Andreev tunneling \cite{SM:Pekola_RMP_2013}, start to affect dynamics of single-electron box (see Eq.~(\ref{SM:rate_eq}) in Supplementary Note 1).
For this purpose we compare the minimal rate of sequential tunneling (see Eq.~(\ref{SM:Gamma[U]}) in Supplementary Note 1 for $U=-E_{\rm C}$) with typical amplitudes of Andreev tunneling rates.
Strictly speaking there is another coherent effect called cotunneling, but it is not relevant to a SEB dynamics due to one tunnel junction in the system.

Substituting the expression for the superconducting density of states $\nu(E)=\left|{\rm Re}\frac{E}{\sqrt{E^2-\Delta_{\rm S}^2}}\right|$ into Eq.~(\ref{SM:Gamma[U]}) in Supplementary Note 1 one can find the expression for the sequential tunneling rate at $k_{\rm B} T\ll U<\Delta_{\rm S}$:
\begin{equation}
\Gamma[U]\approx \frac{\sqrt{2\pi \Delta_{\rm S} k_{\rm B} T}}{e^2 R_{\rm T}}e^{-\Delta_{\rm S}/k_{\rm B} T}\left(1+e^{U/ k_{\rm B} T}\right)
\end{equation}
with the minimal value $\Gamma[-E_{\rm C}]\simeq \sqrt{2\pi \Delta_{\rm S} k_{\rm B} T}e^{-\Delta_{\rm S}/k_{\rm B} T}/{e^2 R_{\rm T}}$ at $E_{\rm C}<\Delta_{\rm S}$.

In subgap region $|U|<\Delta_{\rm S}$ Andreev tunneling rate $\Gamma_{\rm AR}$ can be estimated as follows:
\begin{equation}
\Gamma_{\rm AR} = \frac{R_{\rm Q}}{8 e^2 N R_{\rm T}^2} |U| \ ,
\end{equation}
where $R_{\rm Q} = h/e^2=26$~k$\Omega$ is the resistance quantum, $N$ is the the effective number of channels in the contact.

Using the experimental parameters $R_{\rm T}\sim 12$~G$\Omega$, $|U|\sim\Delta_{\rm S}\sim E_{\rm C}\sim 100$~$\mu$eV, $N\sim 100$ we obtain the following threshold temperature $T^*$ by comparing the rates
$\Gamma[-E_{\rm C}]\sim \Gamma_{\rm AR}$
\begin{equation}
T^* \sim \Delta_{\rm S}/k_{\rm B}\ln\left[8 R_{\rm T} N/R_{\rm Q}\right]\sim 1.15~{\rm K}/\ln[4\cdot 10^8]=60~{\rm mK} \ ,
\end{equation}
below which the coherent effects cannot be neglected.
This estimation agrees very well with the experimental observation of individual Andreev events in a hybrid single-electron transistors.\cite{SM:Maisi_Andr_tun_events}

\section{Supplementary Note 4 Energy gap in the superconductor suppressed by magnetic field}\label{SM_sec:Delta_S(H)}
Experimentally we increase the tunneling rates $\Gamma[U]$ through the NIS junction in the measurements shown in Figs.~3(c) and 4(c) 
 in the main text by applying magnetic field to control the energy gap $\Delta_{\rm S}$ in the quasiparticle spectrum.
Here we estimate the energy gap $\Delta_{\rm S}(H)$ near the junction under the influence of the magnetic field $H = 475$~G applied to the superconducting island made of aluminium.
We focus only on the energy gap value near the junction because the tunneling rates are governed by the local density of states of the superconductor near the junction, as given by Eq.~(\ref{SM:Gamma[U]}) in Supplementary Note 1.

Based on a typical normal-state resistivity $\rho_N=30-40$~n$\Omega\cdot$m of aluminum at $4.2$~K,\cite{SM:2009_Cooler_Timofeev,SM:2010_Thermal_conductance_Peltonen} we estimate the diffusion coefficient $D = 70-90$~cm$^2$s$^{-1}$ from the Drude formula $1/\rho_N = e^2 N_0 D$, where 
$N_0 = 1.45\cdot 10^{47}$~J$^{-1}$m$^{-3}$ is the normal state density of states of aluminium.\cite{SM:2012_Ville_SINIS_overheating}
According to this estimate the elastic mean free path $\ell\sim10$~nm is small-compared to the superconducting coherence length $\xi_0 = \sqrt{\hbar D/\Delta_{\rm S}(0)}\sim 140~-~165$~nm corresponding to the superconducting gap in aluminium at zero magnetic field $\Delta_{\rm S}(0)\simeq 220$~$\mu$eV.
The width of the island near the junction is estimated to be $w = 100$~nm using scanning electron microscope image of the sample.

Due to the inequalities $\ell\ll w\lesssim \xi_0$, one can assume homogeneous suppression of the gap
\begin{equation}
\Delta_{\rm S}(H) = \Delta_{\rm OP}(H) (1-\gamma_H^{2/3})^{3/2} \ ,
\end{equation}
where the expression $\Delta_{\rm OP}(H) = \Delta_{\rm S}(0)\left(1-0.75 \gamma_H-0.54 \gamma_H^2\right)$ for the superconducting order parameter in the magnetic field holds for $\gamma_H\lesssim 0.3$.
These expressions follow from the solution to the Usadel \cite{SM:2003_DOS_in_H} or Gor'kov\cite{SM:1964_Skalsky,SM:1965_Maki} equations. 
Note that in zero magnetic field the order parameter coincides with the energy gap $\Delta_{\rm OP}(0)=\Delta_{\rm S}(0)$.
Here, $\gamma_H = \frac{1}{6}\left(\frac{H \xi_0 w}{\hbar/e}\right)^2$.
This results in the estimate $\Delta_{\rm S}(H) = 96\pm 11$~$\mu$eV for the gap in the presence of the magnetic field applied in the present experiment, which is the value used in the main text, and which agrees with the measured temperature dependence of the tunneling rates (not shown).

\end{document}